\documentclass[aps,prl,twocolumn,showpacs,showkeys]{revtex4-1}

\usepackage{graphicx}
\usepackage{comment}
\usepackage{xcolor}

\usepackage[disable]{todonotes}

\usepackage{epsfig}
\usepackage{color}
\usepackage{setstack}
\usepackage{bm}
\usepackage{amsfonts,amssymb}
\usepackage{mathbbol}
\newcommand{\eqref}[1]{(\ref{#1})}

\newcommand{\im}{{\rm Im}\,}
\newcommand{\re}{{\rm Re}\,}

\newcommand{\hn}{\hat{n}}
\newcommand{\br}{{\bf r}}

\newcommand{\bq}{{\bf q}}

\newcommand{\bK}{{\bf K}}
\newcommand{\bk}{{\bf k}}
\newcommand{\ba}{{\bf a}}
\newcommand{\bb}{{\bf b}}
\newcommand{\bt}{{\bf t}}
\newcommand{\bR}{{\bf R}}
\newcommand{\bM}{{\bf M}}
\renewcommand{\k}{{\bm{k}}}
\newcommand{\q}{{\bm{q}}}
\renewcommand{\v}{{\bm{v}}}
\renewcommand{\r}{{\bm{r}}}

\newcommand{\G}{{\bm{G}}}

\begin{document}

\title{Quasi-flat plasmonic bands in twisted bilayer graphene}

\author{T. Stauber$^{1,*}$ and H. Kohler$^{1,2}$ }

\date{\today}

\affiliation{$^1$Instituto de Ciencias Materiales de Madrid, CSIC,
           C/   Sor Juana In{\'e}s de la Cruz 3, 28049 Madrid,  Spain\\ 
        $^2$Fakult\"at f\"ur Physik, Universit\"at Duisburg-Essen, Lotharstrasse 1,
        47048 Duisburg, Germany}

\begin{abstract}
The charge susceptibility  of twisted bilayer graphene is investigated 
in the Dirac cone, respectively random-phase approximation. For small enough twist angles $\theta\lesssim 2^\circ$ we find weakly Landau damped interband plasmons, i.~e., collective excitonic modes which exist in the undoped material, with an almost constant energy dispersion. In this regime, the loss function can be described as a Fano resonance and we argue that these excitations arise from the interaction of quasi-localised states with the incident light field. These predictions can be tested by nano-infrared imaging and possible applications include a "perfect" lens without the need of left-handed materials.
\end{abstract}
\email{tobias.stauber@csic.es}
\maketitle

{\em Introduction.} With the discovery of graphene and other two-dimensional (2D) crystals \cite{Novoselov05}, the field of plasmonics has received renewed attention \cite{Koppens11,Grigorenko12,Stauber14,Abajo14,Low14,Peres16}. Especially single-layer graphene on hexagonal boron-nitride (h-BN) displays outstanding properties, hosting long-lived plasmons with life times of the order of 500$fs$ and offering the possibility of tuning the plasmonic resonances via an electrostatic gate \cite{Woessner15}. The plasmonic modes can also be modified when the 2D materials form Moir\'e patterns with the underlying substrate and emerging high-energy modes were observed for graphene on top of h-BN \cite{tom14,Ni15}. Along these lines, twisted bilayer graphene offers new perspectives for tuning the electromagnetic response by changing the twist angle \cite{bri12,hav14,schmi14,pat15,Yin16}. Here, we will investigate the plasmonic spectrum of twisted bilayer graphene using the continuous model by Lopes-Santos et. al. \cite{lop07}, i.e., we extend previous results for the local conductivity to finite momentum transfer \cite{moo13,sta13}. For small enough twist angles $\theta\lesssim 2^\circ$ we will find novel weakly Landau damped interband plasmons, i.~e., collective excitonic modes which exist in the undoped material with an almost constant energy dispersion, arising from quasi-localised states.

Plasmons are collective charge oscillations leading to nanoscale optical fields and thus they are linked to the existence of a plasma, i.e., to a finite charge stiffness or Drude weight, $D$. Within a hydrodynamic model, the plasmon energy is related to the Drude weight as $\omega_p\sim \sqrt{D}$ which rules out the existence of plasmons for neutral systems for which $D=0$. Nevertheless, here we will show that for sufficiently small twist angle close to the magic angle at which the Fermi velocity becomes zero and flat bands develop \cite{bis11}, genuine collective modes emerge even in the case of zero chemical potential. These excitations prevail for not too large finite chemical potential and can, therefore, be interpreted as interband plasmons or  collective excitonic oscillations.

As argued above, terahertz plasmons in graphene and/or superlattices are only present at finite chemical \cite{Shung86,Hawrylak85}. But so-called $\pi$-plasmons can also be observed in neutral free-standing graphene at energies $\omega_p^\pi\gtrsim4.5$eV \cite{Eberlein08,Kinyanjui12}. These are related to a van Hove singularity \cite{sta10}, and an obvious guess would be that there will be similar plasmonic excitations at lower energies due to the appearance of emerging van Hove singularities located between the two Dirac cones of the two twisted layers \cite{Li10,Luican11}. However, the above plasmons are invoked by delocalised $\pi$-electrons, whereas the plasmons discussed here originate from quasi-localised states, reminiscent to a recent study on localised plasmons in disordered graphene \cite{Muniz10} and bilayer nano-disks \cite{Wang16}.

The emergence of interband plasmonic modes around the neutrality point is related to the deviation from Dirac fermion behavior in the charge response $\chi$, i.e., the imaginary part Im$\chi$ has to decay faster than $\omega^{-1}$ for $\omega\to\infty$. Interband (out-of-phase) plasmons with a linear (sound-like) dispersion should therefore be hosted by topological insulators such as mercury telluride described by the BHZ-model which mixes Dirac with Schr\"odinger electrons \cite{Juergens14a,Juergens14b}. In the case of twisted bilayer graphene, we also find deviations from the typical Dirac response for small twist angles, however, here the emerging interband plasmons arise through the interaction of the incidence light with quasi-localised states, displaying an almost constant dispersion with energies $\hbar\omega\sim20-200$meV, tuneable by the twist angle. Furthermore, they carry a dipole moment (in-phase plasmons) and it should, therefore, be easier to observe genuine interband plasmons in twisted bilayer graphene with twist angles of $\theta\lesssim 2^\circ$ via, e.g., nano-infrared imaging or scattering-type scanning near-field optical microscopy (s-SNOM) \cite{Chen12,Fei12}. 

Before we outline the explicit calculations, let us specify our definition of plasmonic excitations. Often, a peak in the electron energy loss function serves as criterion for a plasmon mode. Nevertheless, this is only an indication for an enhanced charge response and not for collective oscillations which are indicated by a pole in the two-particle Green's function or, alternatively, a zero in the (real part of) the dielectric function. Within this definition, the modes found in Refs. \cite{Liu08,Tegenkamp11,Politano11,Langer11} for undoped graphene or energies larger than twice the chemical potential on various substrates are not genuine plasmons as discussed in Ref. \cite{Stauber14}.

\begin{figure*}[t]
\includegraphics[width=2\columnwidth]{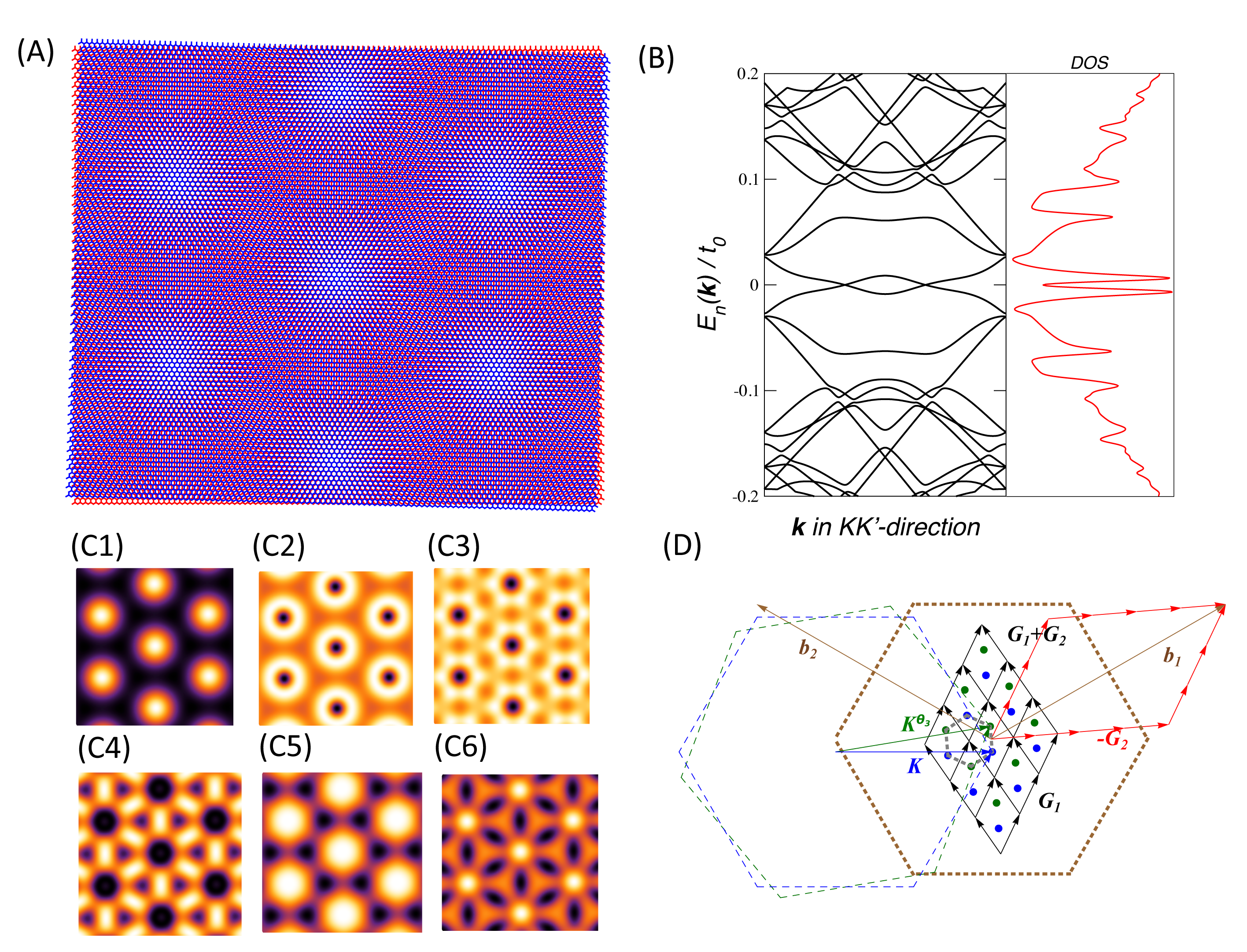}
\caption{\label{Fig_TBG} (Online color)  (A) Real space image of twisted bilayer: Regions of AA-stacked graphene are surrounded by regions of AB-stacked graphene which form the Mori\'e-pattern. (B) Band-structure with corresponding density of states and (C1)-(C6) the local density corresponding to the first 6 conduction bands (CB). Clearly seen are the localised states inside the AA-stacked regions, especially in the case of the lowest CB. In all cases, the twist angle corresponds to $i=20$ ($\theta_{20}=1.61^\circ$). (D) Sketch of the Brillouin zone of twisted bilayer for a large angle with $i=3$, $\theta_3=9.14^{\circ}$. Details are given in the Supplementary Information.}
\end{figure*}

{\it Model.} 
In order to describe twisted bilayer graphene, we follow Refs.~\cite{lop07,mel10} with the intra(inter)layer hopping amplitude $t_0=2.78$($t_\perp=-0.33$)eV. A self-contained discussion on the model is given in the Supplementary Information (SI) \cite{supplmaterial}. In the following, we focus on a discrete set of twist angles $\theta_i$ which are labelled by only one integer $i$ through $\cos(\theta_i) \ =\ 1-\frac{1}{2A_i}$ with $A_i =  3i^2+3i+1$. The unit cell of twisted bilayer graphene is $A_i$ times larger than the unit cell of single or bilayer graphene and the unit vector thus of size $\sqrt{A_i}a$ with $a\approx 2.46\AA$. In Fig.~\ref{Fig_TBG}, (A) the real-space image, (B) the band-structure, and (C$n$) the local density corresponding to the $n$-th conduction band are shown for an twist angle $\theta_{i=20}=1.61^\circ$, as well as the extended Brillouin zone for $\theta_{i=3}=9.14^\circ$ (D). 

For large angles, the spectrum can be approximately described by a single parameter $\alpha=\frac{\sqrt{A_i/3}}{2\pi}\frac{t_\perp}{t_0}$ and perturbation theory correctly predicts the emerging Dirac cone physics with renormalised Fermi velocity given by $v^*_{\rm F}$ $\approx$ $v_{\rm F}(1-3\alpha^2)/(1+6\alpha^2)$ with $ \hbar v_F=\sqrt{3}at_0/2$ \cite{lop07,bis11}. But for small twist angles with $\theta<2^\circ$, a new regime occurs where the bands are very flat and the velocity tends to zero \cite{Trambly10}, see Fig.~\ref{Fig_TBG} (C1). It is this new regime that will give rise to novel interband plasmons.

 {\it Dynamical charge response.} 
The dynamical density-density response function is defined as $\chi(\br,\br^\prime,t)\ =\ \frac{1}{i\hbar}\Theta(t)\left< \left[\hn(\br,t), \hn(\br^\prime)\right]\right>$ and for $|\bq|\ll|\G_i|$ its Fourier transform in space and time, $\chi(\bq,\omega)$, is well-defined. Let $E_m^\tau(\bk)$ and $|\tau,m,\bk\rangle$ denote eigenvalues and eigenfunctions
of the effective Hamiltonian given in the SI \cite{supplmaterial}. Then  $\chi(\bq,\omega)$ can be expressed in the long wavelength limit as 
\begin{eqnarray}
\label{corr}
 \chi(\bq,\omega) & = &  \frac{ g_s}{V}\sum_{\tau=\pm} \sum_{m,n} \sum_{\bk\in1.BZ} |\langle \tau,n,\bk+\bq |\tau,m,\bk\rangle |^2\nonumber\\
 &\times&\frac{n_F[E_m^\tau(\bk)]
 - n_F[E_n^\tau(\bk+\bq)]}{\hbar\omega- E_n^\tau(\bk+\bq)+E_m^\tau(\bk)+i0}\;.
\end{eqnarray}
Here, $n_F[x]=  (e^{\beta(x-\mu)}+1)^{-1}$ is the Fermi function and $g_s=2$ the spin degeneracy. The sum over $\bk$ is over the first Brillouin zone of the supercell and $m,n$ denote the band-indices. Note that Eq. (\ref{corr}) comprises eigenvalues and eigenstates of both inequivalent Dirac points $\tau$, such 
that  $\chi(\bq,\omega)$ manifestly fulfils  the usual symmetry relations of a response function \cite{giu05}, namely  $\chi(\bq,\omega)=\chi^*(\bq,-\omega)$ and $\chi(\bq,\omega)=\chi( -\bq,-\omega)$ due to time reversal invariance and a real response, respectively.   

For energies $\hbar\omega\gg t_\perp$, the effects of the interlayer coupling become negligible and the result for two  
decoupled graphene monolayer at zero temperature and zero chemical potential \cite{gon94}
\begin{equation}
\label{chimono}
\chi_{0}(\bq,\omega) \ = \  \frac{-i g_{\ell} g_{v} g_{s}}{16 \hbar}\frac{q^2}{\sqrt{\omega^2-(v_{\rm F}q)^2}}
\end{equation}
must be recovered ($g_{\ell} =2$ and $g_{v}=2$ are the layer and the valley degeneracy). This result holds for vanishing coupling strength $t_\perp$ or, more precisely, for vanishing $\alpha$. 

\begin{figure*}[t]
\includegraphics[width=2\columnwidth]{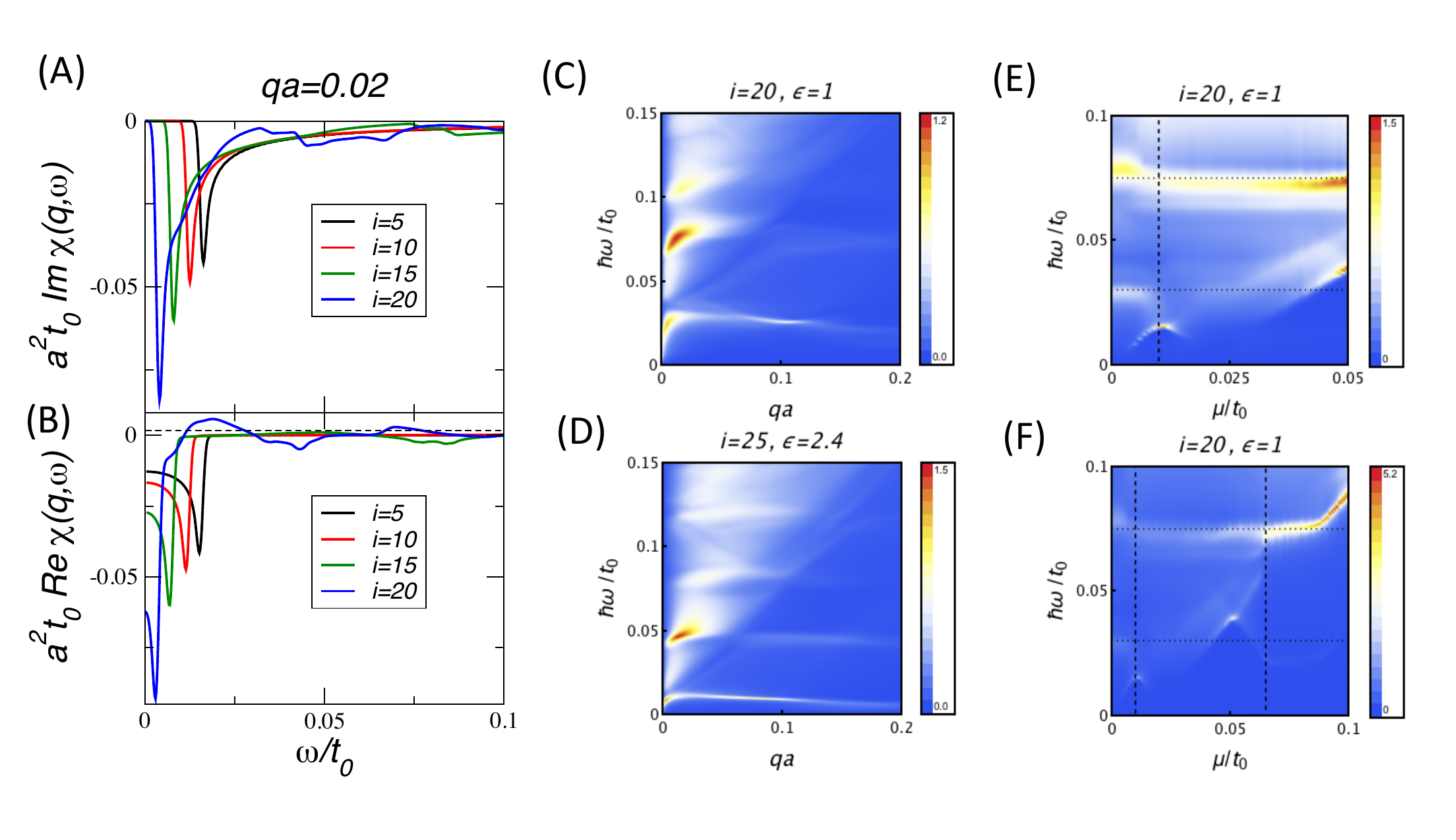}
\caption{\label{Fig_NR} (Online color)  Imaginary (A) and real (B) part of the charge response $\chi(\bq,\omega)$ for different twist angles $\theta_{5}\approx 6.01^\circ$ (black), $\theta_{10}\approx 3.14^\circ$ (red), $\theta_{15}\approx 2.13^\circ$ (green) and $\theta_{20}\approx 1.61^\circ$ (blue). The in-plane momentum transfer is $qa=0.02$ in the direction of $\Delta\bK$ and the dashed lines indicate the zero of the real part of Eq. (\ref{DielecticFunction}) for an effective dielectric medium with $\epsilon=1$. Density plot for the loss function $S(\bq,\omega)=-\im\epsilon^{-1}(\bq,\omega)$ of twisted bilayer graphene with a twist angle $\theta_{20}\approx1.61^\circ$ with $\epsilon=1$ (C) and $\theta_{25}\approx 1.30^\circ$ with $\epsilon=2.4$ (D) as a function of frequency $\omega$ and in-plane momentum $q$. In panels (E) and (F): Density plot for the loss function $S=-\im\epsilon^{-1}(\bq,\omega)$ of twisted bilayer graphene with a twist angle $\theta_{20}\approx1.61^\circ$ at fixed wave number $qa=0.02$ and $\epsilon=1$ as function of frequency $\omega$ and chemical potential $\mu$ for two different scales. Also shown vertical dashed lines at $\mu/t_0=0.01,0.065$ and horizontal dotted lines at $\hbar\omega/t_0=0.03,0.075$ as guide for the eye.}
\end{figure*}

For small $\alpha$, the renormalization of the Fermi velocity is expected to be the main effect and the response function should be well described by 
$\chi^*_{0}(\bq,\omega)$, which is defined as  $\chi_{0}(\bq,\omega)$ in Eq.~(\ref{chimono}) but with $v_{\rm F}$ replaced by  $v^*_{\rm F}$.   
Obviously the  locus of the singularity of $\chi^*$ moves towards zero as the Fermi velocity decreases while its spectral weight  increases with $1/\sqrt{v_{\rm F}^*}$.  This can be seen in Fig.~\ref{Fig_NR} (A) for large twist angles with $i=5,10$.

{\it Static response.} 
For large twist angle, the electronic spectrum of twisted bilayer graphene is characterised by two Dirac cones per valley and the static susceptibility thus scales linear with the wave-number, i.e., Re$\chi(\bq,0)\sim|\bq|$ as seen in Eq. (\ref{chimono}). Remarkably, for small twist angles $\theta\lesssim\theta_{i=20}$, the static response becomes quadratic for small momenta, i.e., $\chi(\bq,0)\sim |\bq|^2$, see SI \cite{supplmaterial}. This departure from Dirac cone physics is crucial to host genuine plasmons at zero doping \cite{Juergens14a,Juergens14b}. 

{\it Numerical results.} In Fig.~\ref{Fig_NR} (A),  the imaginary part of $\chi(\bq,\omega)$ is shown for various angles at constant momentum transfer $qa=0.02$ in the direction of $\Delta\bK=\bK-\bK^{\theta}$. With $\omega^*$ denoting the frequency at which Im$\chi$ becomes maximal, $\im\chi(\bq,\omega^*)\sqrt{\omega^*}$ is independent of $i$ for $i\lesssim15$ consistent with a square-root divergency. Furthermore, we have $\omega^*=v_F^*q$ for $i\lesssim15$ whereas for smaller angles with $i\gtrsim15$, $\omega^*>v_F^*q$ and $\im\chi(\bq,\omega^*)\sqrt{\omega^*}$ is not constant anymore. We thus observe a departure from Dirac cone physics for small angles in the charge response.  

The crossover behaviour around $i\approx15$ can also be observed in the real part of the susceptibility, Fig.~\ref{Fig_NR} (B). For $i \gtrsim15$, Re$\chi$ becomes positive for certain frequencies which opens up the possibility for the existence of genuine plasmons with energy $\hbar\omega_p$. This is indicated by the dashed line where the dielectric function within the random-phase approximation (RPA) becomes zero,
\begin{equation}
\label{DielecticFunction}
\epsilon(\bq,\omega_p)=1-v_q\chi(\bq,\omega_p)=0\;.
\end{equation}
Above, we defined the Coulomb potential $v_q=\frac{e^2}{2\epsilon_0\epsilon|\bq|}$ for an effective dielectric medium with static dielectric constant $\epsilon$. This has been set equal to one (vacuum) in the corresponding dashed curve and we will also discuss results for a finite value with $\epsilon=2.4$, corresponding to a twisted bilayer graphene on top of SiO${}_2$
.

{\it Loss function.} Undamped plasmons only exist if $\epsilon(\bq,\omega)$ $=$ $0$.  Nevertheless, plasmons  with frequency $\omega_p$ can also be defined for ${\rm Im}\chi\neq 0$  by the condition  ${\rm Re}\, \epsilon(\bq,\omega_p)= 0$ as long as the loss function $S(\bq,\omega)=-{\rm Im} \varepsilon^{-1}(\bq,\omega)$ is peaked around $\omega_p$ with width $\gamma\ll\omega_p$. This condition allows the plasmon to oscillate sufficiently long before decaying through Landau damping into the particle--hole continuum and renders it detectable by e.g. nano-infrared imaging \cite{Chen12,Fei12}.
 

In Fig.~\ref{Fig_NR} (C), the density plot of the loss function for a surrounding medium with $\epsilon=1$ is shown as a function of momentum and of frequency for a twist angle $\theta_{20}\approx 1.61^\circ$. The in-plane momentum vector points into the direction of $\Delta\bK$ but the results hardly depend on the direction of $\bq$. In Fig.~\ref{Fig_NR} (D), the loss function is shown for $\theta_{25}\approx 1.30^\circ$ with $\epsilon=2.4$. For this angle, even with an effective dielectric medium up to $\epsilon=4$, genuine plasmons with Re$\epsilon(\bq,\omega_p)=0$ are present at $T=0$. For both angles, we observe several almost equally spaced plasmon branches extending to large $q$-values. These quasi-localised plasmons emerge from quasi-localised eigenstates as we will argue below and are the main observation of this work.

For smaller momentum transfer $qa\approx0.02$, the two lowest resonances show an asymmetric line shape which can be well fitted by a Fano resonance, see SI \cite{supplmaterial}. This is because the loss function is directly related to the extinction spectrum for which Fano resonances are well-known for confined plasmonic systems and which occur when localised states interact with a continuum \cite{Giannini11}. The asymmetry increases with decreasing angle and also the peak position shifts to lower energies, see SI \cite{supplmaterial}. These results go in line with the stronger localisation for twist angles close to the magic angle at $\theta_{i=31}\approx1.05^\circ$ as well as the increasing dot size given by the AA-stacked island proportional to $\sqrt{A_i}\propto i$. 

{\it Local field effects.} For small angles and/or large wave number, local field effects have to be taken into account since the wave number $q$ becomes comparable to the length of the first reciprocal lattice vector $|\G_i|=|\G_0|/\sqrt{A_i/3}$ with $|\G_0|a=4\pi/\sqrt{3}$. In the SI, we analyse the local field effects on the loss function and conclude that there are no significant changes, i.e., the plasmonic resonances are only slightly shifted but prevail \cite{supplmaterial}.

{\it Finite doping, interlayer bias, disorder and temperature.} For finite chemical potential with $\mu\lesssim\hbar\omega_p^m$, $m$ counting the plasmonic resonances, the peaks with energy $\omega_p^m$ prevail. This supports our interpretation of the collective excitonic excitations due to interband transitions, i.e., novel interband plasmons due to the hybridisation of the localised states with the incident light field. Plasmons can thus be tuned and quenched/enhanced by changing the twist angle and chemical potential, respectively, as seen in Fig. \ref{Fig_NR} (E) and (F) which show the loss function as function of frequency and chemical potential at fixed in-plane momentum $qa=0.02$ for two different scales.

By applying an interlayer bias $\Delta$, a gap is opened in the spectrum of Bernal(AB)-stacked graphene bilayer \cite{McCann06}, but for twisted bilayer graphene only the energy levels of the two Dirac points of one valley are shifted to positive and negative energies, respectively. Again, the localised plasmon modes are preserved for $\Delta\lesssim\hbar\omega_p^m$, showing the robustness of these collective excitonic oscillations. In fact, also for larger interlayer bias $\Delta\gtrsim\hbar\omega_p^m$, the resonances persist supported by the {\em local} gap present in the AB-stacked regions, see SI \cite{supplmaterial}.

Disorder can be qualitatively modelled by introducing a finite damping term in Eq. (\ref{corr}). Numerically, we first obtain Im$\chi$ and then Re$\chi$, invoking the Kramers-Kronig relation. A moderate broadening in Im$\chi$ and consequently in Re$\chi$ does not alter our general predictions. The same holds for finite room-temperature, see SI \cite{supplmaterial}.

 {\it Real space interpretation.} The novel plasmon modes consist of collective interband transitions and therefore, the corresponding electron and hole densities are equal. In an extended systems, electron and hole densities must move out of phase in order to generate a restoring force which maintains the charge oscillations. If the system is partially confined due to an external potential, electron and hole densities can also move in-phase making them susceptible to dipole coupling to an incident light field. 

Assuming the confinement to be harmonic, the spectrum is given by equally spaced energy levels. Moreover, the center-of-mass equation of motion is linear and all Fourier-components move with the same frequency, i.e., the dispersion is constant and independent of $q$. Both features are reflected by the loss function of twisted bilayer graphene which has to be contrasted to the case of interlayer plasmons in mercury telluride which shows a linear dispersion in accordance to out-of-phase oscillations which do not couple to light \cite{Juergens14a,Juergens14b}. A simple model describing these quasi-confined regions is discussed in the SI \cite{supplmaterial}.

Exciting the system by s-SNOM, a particle-hole or excitonic density is created, oscillating within several adjacent AA-stacked regions of quasi-localized wave functions. Moreover, there is a linear shift in the resonant plasmon energy for different twist angles with $\omega_p^m\sim1/R$, where $R\sim\sqrt{A_i}\sim i$ denotes the radius of the localised AA-stacked region which is approximately linear for large $i\gtrsim15$, see SI \cite{supplmaterial}.  


{\it Applications.} A plasmonic resonance with almost constant dispersion at $\hbar\omega_0$ opens up several possible applications. Let us highlight here a device with two twisted bilayer graphene layers on top and on the bottom of a dielectric $\epsilon$ of width $d$. Following Ref. \cite{Stauber12}, we find exponential amplification of the near-field modes at constant energies $\omega_{exp,1}=\omega_0+O(e^{-2qd})$ and  $\omega_{exp,2}=\frac{\epsilon-1}{\epsilon+1}\omega_0+O(e^{-2qd})$. A "perfect" lens in the spirit of Pendry \cite{Pendry00} can thus be designed without the need of left-handed materials. 

Also extraordinary absorption of propagating light at $\hbar\omega_0$ is expected due to coupling to the reciprocal vector of the Moir\'e-superlattice. For a polarization in direction of $\G_1$ and twist angle $\theta_{i=25}$, we have $|\G_1|a\approx0.16$ and peaks in the loss function correspond to enhanced absorption.

{\it Summary and Discussion.}  
We have predicted novel interband-plasmons in undoped twisted bilayer graphene for small twist angles. Moreover, we showed that the plasmonic excitations are connected to the deviations of Dirac cone physics and consequently to quasi-localised states giving rise to Fano resonances. This makes twisted bilayer graphene an exciting new metamaterial with extraordinary properties leading to enhanced absorption and exponential amplification at constant energy giving rise to the possibility of a "perfect" lens without the need of left-handed materials.

The new interband plasmonic modes can be interpreted as collective excitonic in-phase oscillations in a periodic, but quasi-confining potential surrounding the AA-stacked regions. We thus expect these modes to also emerge in other systems with electronic (quasi-)confinement and/or commensurate structure. 

{\it Acknowledgements}. The authors thank Luis Brey and T.S. Guillermo G\'omez-Santos. Support by Grants FIS2014-57432-P, S2013/MIT-3007 MAD2D-CM.

{\Large\bf Supplementary Information} 
\section*{The continuous model for twisted bilayer graphene}
While twisted bilayers were often addressed theoretically by first principle calculations \cite{sua10,Trambly10, hav14, jun14,miw15}, a continuos model based on the Dirac cone approximation of the tight binding Hamiltonian 
was proposed by Lopes dos Santos et al. \cite{lop07,lop12}. Also related models were developed and employed in Refs. \cite{bor09, bis11,moo13,wal13}. Albeit similar, the models differ in certain details, e.g., in Ref. \cite{moo13} particle hole symmetry is conserved, while in Ref. \cite{lop07}, it is broken.

In twisted graphene bilayers, Moir\'e patterns may appear which exhibit the same hexagonal lattice structure as in single layer graphene. Depending on the twist 
angle between the two layers, the length of the lattice vectors of this superstructure might be largely enhanced as compared to  $a\approx 2.42$A, the length of the lattice vectors   
$\ba_{1,2} = a\left(\pm1/2, \sqrt{3}/2\right)$ of graphene monolayer. Strictly speaking, not all twist 
angles are allowed but only a commensurate set $\theta_{mn}$ which  map 
the lattice point $(m \ba_1 ,n\ba_2)$ 
onto $(n \ba_1 ,m\ba_2)$, $m,n\in \mathbb{N}$ \cite{cam07}. However, the set of possible angles is dense in 
$[0,2\pi]$ such that the spectral properties at low energies continuously depend only on
the value of  the angle and not on the integers  $n,m$. 

\begin{figure}
\begin{center}
\includegraphics[scale=1]{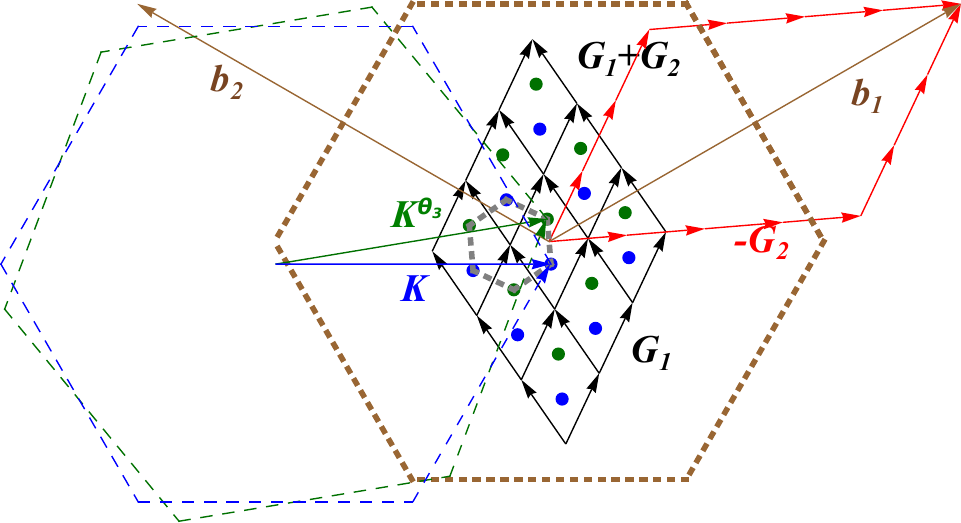}
\end{center}
\caption{\label{Fig1} (Coloronline): Sketch of the twisted bilayer in momentum space for a large angle $i=3$. The Brioullin zones of the two layers (dashed green and blue) are rotated by an angle $\theta_3\approx 9.14^{\circ}$. The Dirac points $\bK$ and $\bK^{\theta_i}$ are the two inequivalent Dirac points of the Brioullin zone of the superlattice (small dashed grey hexagon). The lattice vectors ${\bf b}_{1,2}$ of the reciprocal lattice of graphene monolayer (brown arrows) are related to the lattice vectors of the reciprocal superlattice (red arrows) by $ {\bf b}_1= i\G_1-(i+1)\G_2$.} 
\end{figure}

In this work, we focus on  $m-n=1$,  obtaining a discrete set of twist angles $\theta_i$ which are labelled by only one integer $i$ through 
\begin{equation}
\cos(\theta_i) \ =\ 1-\frac{1}{2A_i} \ , \qquad A_i =  3i^2+3i+1\ .
\end{equation}
An arbitrary small twist angle can be achieved by increasing the integer $i$ keeping in mind that all 
intermediate angles can be approximated to arbitrary accuracy by allowing 
two integers.  The lattice vectors of the superlattice are 
\begin{equation}
\bt_1 = i \ba_1+ (i+1) \ba_2 \ ,  \quad \bt_2  =  -(i+1) \ba_1+ (2i+1)\ba_2
\end{equation}
spanning a super unit cell with an area $A_i$ times large than the unit cell of single layer 
graphene. 
Likewise, the area of the reciprocal superlattice spanned by the vectors
\begin{eqnarray}
\G_1 & = & \frac{1}{A_i}\Big( (2 i+1) \bb_1+ (i+1)\bb_2\Big) \ , \nonumber\\
\G_2 & = & \frac{1}{A_i}\Big( -(i+1) \bb_1+ i \bb_2\Big) \ ,
\end{eqnarray}
is $A_i$ times smaller than the area of the reciprocal lattice of graphene, spanned by $\bb_{1,2}
=\frac{2\pi}{a}(\pm1,1/\sqrt{3})$. The Brioullin zones of the two monolayers are twisted by the angle 
$\theta_i$, such that the Dirac points $\bK^\phi$, $\phi= 0,\theta_i$, ($\bK\equiv \bK^{\phi=0}$) at the positions 
$\bK^\phi= (4\pi/3a)(\cos\phi,\sin\phi)$ are connected by the vector $\Delta\bK$ $\equiv$ $\bK^{\theta_i}-\bK $ $=$ $\left( 2\G_1+\G_2\right)/3$, 
which decreases with decreasing twist angle $\theta_i$ as $|\Delta\bK|$ $= |\bK|/\sqrt{A_i}$. The two points $\bK$ and $\bK^{\theta_i}$ are the vertices of the Brioullin zone of the reciprocal superlattice with $\Gamma$--point at $\bK^{\theta_i}+(\G_1+\G_2)/3$ and the central $M$--point at ${\bf M}$ $ = $$(\bK+\bK^{\theta_i})/2$. 
The geometry in momentum space is summarized in Fig. \ref{Fig1}.

We follow Refs.~\cite{lop07,mel10} and take into account 
interlayer hopping only between the two sites in each layer 
which are closest to each other. The hopping amplitude $t(\bR,\boldsymbol \delta)$ between a lattice site $\bR$ on 
the first layer and site $\bR+\boldsymbol \delta+d{\bf e}_3$ of the second one is much smaller than the intralayer hopping 
amplitude $t_0\approx 2.78$ eV. In principle  $t(\bR,\boldsymbol \delta)$ has a rather complicated space dependence, 
but its periodicity with respect to the 
superlattice allows an expansion in Fourier components $\tilde{t}_{\sigma \sigma^\prime}(\G)$. 
These depend on the reciprocal vectors of the superlattice $\G$  and on 
the sublattices of the two layers.  

Using the Dirac cone approximation for single layer graphene, the Hamiltonian for twisted bilayer graphene close to the Dirac point $\bK$ reads $\hat{H}=$$\sum_{\bk}\hat{H}(\bk)$ with \cite{lop12,bis11}
\begin{eqnarray}
\hat{H}(\bk)&=&  \hbar v_{\rm F} \sum_{\phi=0,\theta_i} \hat{a}^\dagger_{\phi A  \bk} \hat{\bm\sigma}^\phi_{\rm AB}\cdot(\bk+{\bf M}-\bK^\phi) \hat{a}_{\phi B \bk}\nonumber\\\label{ham1}
&+& \sum_{\sigma,\sigma^\prime,\G}\hat{a}^\dagger_{0 \sigma  \bk+\G}   \tilde{t}_{\sigma \sigma^\prime}(\G)  \hat{a}_{\theta_i \sigma^\prime \bk} +{\rm h.c} \ .
\end{eqnarray} 
Here,  $\hat{a}^\dagger_{\phi\sigma \bk}$ creates an electron on a single layer with twist angle $\phi$ on sublattice $\sigma=A,B$ with lattice momentum $\bk+\bM-\bK^\phi$. Moreover,  $\hat{\bm \sigma}^\phi$ $=$ $e^{i\phi\hat{\sigma}_z/2}\hat{\bm \sigma}e^{-i\phi\hat{\sigma}_z/2}$ with $\hat{\bm \sigma}$ $= (\hat{\sigma}_x,\hat{\sigma}_y)$ and the Pauli matrices $\hat\sigma_x$, $\hat\sigma_y$. The Fermi velocity of a graphene monolayer is $v_{\rm F} = \sqrt{3}at_0/(2\hbar)$$\approx$ $9\times 10^5$ms$^{-1}$. The Dirac cone approximation is valid for $\bk$-values much smaller than $|\bK|$ $=$ $4\pi/(3a)$, respectively for energies  much smaller than $\hbar v_{\rm F}|\bK|$ $\approx$ $10.1$ eV. In practice we will consider frequencies in a  range $\omega <\Lambda$, where $\Lambda\sim 0.8 t_0/\hbar$ corresponding to a cutoff wave vector $ k_{\Lambda}\approx a^{-1}$.

The interlayer hopping was investigated in detail in \cite{lop07,lop12} and with a different but equivalent approach in \cite{bis11}. In Ref. \cite{lop07} it was pointed out that the  modulus of the interlayer hopping matrix element is independent of the sublattice index 
$t_\perp(\G)=|\tilde{t}_{\sigma\sigma^\prime}(\G)|$. Moreover  $t_\perp(\G)$ decreases algebraically with $aA_i | \Delta\bK + \G|$. Thus it is well justified to consider only  these values  for $n,m \in\mathbb{Z}$, $ \G = n\G_1+m\G_2$  for which this quantity is minimal and neglect all others. There exist three pairs of integers
$(n,m)=$ $(0,0),(-1,0),(-1,-1)$  yielding the same minimal value $t_\perp = t_\perp(0)$. Note the asymmetry in these pairs of integers, which breaks translation invariance in $\bk$--space. The phases of  $\tilde{t}_{\sigma\sigma^\prime}(\G)$ were worked out in \cite{lop07,bis11} utilizing geometric arguments ($ \xi= 2\pi/3$)
 \begin{equation}
 \tilde{t}(0)\ = \  t_\perp\left(\begin{array}{cc} 1&1\cr 1&1 \end{array}\right)\ ,\  \tilde{t}(-\G_1)\ = \  t_\perp\left(\begin{array}{cc} e^{i\xi}& 1\cr e^{-i\xi}& e^{i\xi} \end{array}\right)\ ,
 \end{equation}
and $\tilde{t}(-\G_1-\G_2)= \tilde{t}^*(-\G_1)$. 

We choose for the interlayer hopping strength the value $t_\perp=-330$ meV analogous to Refs. \cite{bis11, sta13}.  However experiments are not conclusive on the exact value of $t_\perp$, which seems to depend on the method used to synthesise the sample. Note that  the Hamiltonian (\ref{ham1}) is only a valid approximation close to the  Dirac point $\bf K$ and breaks time reversal invariance (TRI).  In order to restore TRI, one has to consider  both inequivalent Dirac points. Let $\hat{H}^{\prime}(\bk)$ denote the Hamiltonian close to the second Dirac point ${\bf K}^{\prime}$ $=$  $-{\bf K}$. 
We then find $\hat{H}(\bk)$$=$ $(\hat{H}^\prime(-\bk))^T$ and therefore TRI is conserved in the full Hamiltonian. Thus, it suffices to focus on a single Dirac point. Note that the Hamiltonian (\ref{ham1}) breaks particle--hole symmetry in contrast to the one employed by Moon and Koshino \cite{moo13}.
 
 For small angles, the rotated spin operator $\hat{\bm \sigma}^{\theta_i}$ can be approximated by the unrotated one $\hat{\bm \sigma}^{\theta_i}$  $\approx$ $\hat{\bm \sigma}$. Within this approximation the spectrum of the 
 Hamiltonian of Eq. (\ref{ham1}) only depends on the single parameter $\alpha= \sqrt{A_i}t_\perp/(3\hbar v_{\rm F}|\bK|)$.  However, the full Hamiltonian $\hat{H}$ comprises all momenta within an area $\sim \pi k_\Lambda^2$ which is covered by approximately $N\approx A_i/14$ unit cells of the reciprocal superlattice, which give rise to the same amount of Moir\'e bands.  
 
\section*{Static response}
The susceptibility is related to the overlap between the wave function at two momenta which differ by a fixed momentum $\q$, i.e., $|\langle \tau,n,\bk+\bq |\tau,m,\bk\rangle |^2$, see Eq. (1) of the main text. Expanding the eigenstate for small $\q$, i.e., $\langle \tau,n,\bk+\bq|=\langle \tau,n,\bk|+\bq\cdot\partial_\k\langle \tau,n,\bk|$, will lead to a static susceptibility quadratic in $q$, $\re\chi\sim q^2$, due to the orthogonality of states at the same momentum $\k$ for $n\neq m$. 

For Dirac Fermions, this quadratic behaviour is changed to become linear, $\re\chi\sim q$, as also seen from the general result of Eq. (2) of the main text. This is due to the linear dispersion which leads to an energy denominator in Eq. (1) of the main text proportional to $q$ and thus a cancellation. 

\begin{figure}[t]
\includegraphics[width=0.99\linewidth]{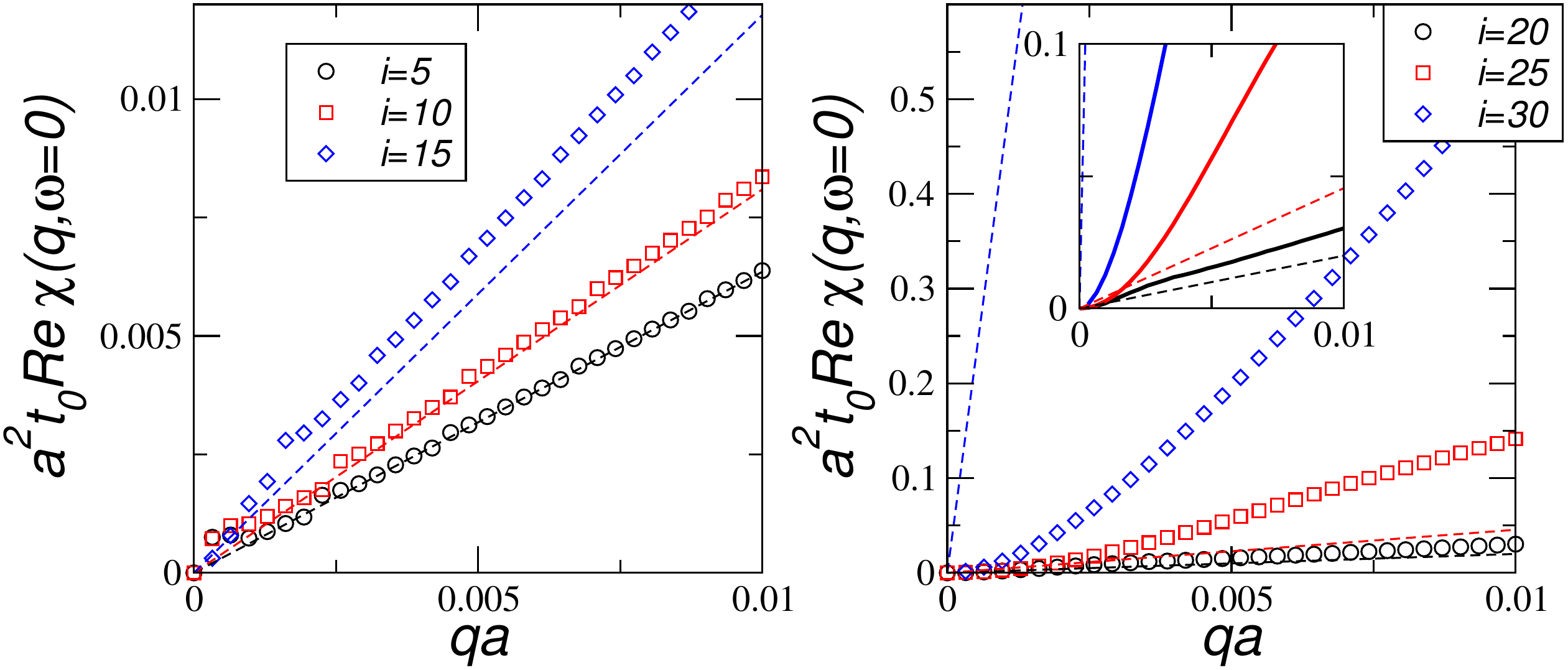}
\caption{\label{Static} (Online color) Static susceptibility Re$\chi$ for different angles $\theta_i$ corresponding to $i=5,10,15$ (left) and $i=20,25,30$ (right). Also shown the static response of Dirac Fermions with renormalised velocity $v^*_{\rm F}$ (dashed lines).}
\end{figure}

In Fig. \ref{Static}, the static susceptibility for different angles $\theta_i$ is shown corresponding to $i=5,10,15$ (left) and $i=20,25,30$ (right). Also shown is the static response of Dirac Fermions with renormalized Fermi velocity $v_{\rm F}^*$ which is given by $v^*_{\rm F}$ $\approx$ $v_{\rm F}(1-3\alpha^2)/(1+6\alpha^2)$ with $\hbar v_F=\sqrt{3}at_0/2$.\cite{bis11} This yields a good approximations for large angles shown on the left panel. On the right panel, the renormalised Fermi velocity is either too large ($i=20,25$) or too small ($i=30$). For $i=30$, e.g., $v_{\rm F}^*=0.001v_F$ which is already close to the magic angle with $i=31$ for which $v_{\rm F}^*\approx0$. A better approach would be to obtain the renormalised Fermi velocity from the maximum in Im$\chi(\q,\omega^*)$ with $\omega^*=v_F^*q$. This yields better results for small $q$-values, but again fails to predict the general behaviour for large momenta.


Clearly seen on the right panel is the deviation from linear behaviour of Re$\chi$ for small $q$-values, see also the inset. In fact, for $i=30$ a clear quadratic behaviour extends over a wide range of momenta indicating the break-down of the linear Dirac-cone physics. But also for $i=25$ and even for $i=20$, a quadratic dispersion sets in for small $q$-values. The break-down of Dirac-cone physics is crucial to obtain the energetically low-lying plasmonic modes at zero doping as discussed in Refs. \cite{Juergens14a,Juergens14b}.

\section*{Angle dependence of the loss function}
Here, we will discuss the loss function for a wider parameter regime and also determine the fitting parameters for the Lorentzian and Fano resonance, respectively.

\begin{figure}[t]
\includegraphics[width=0.99\linewidth]{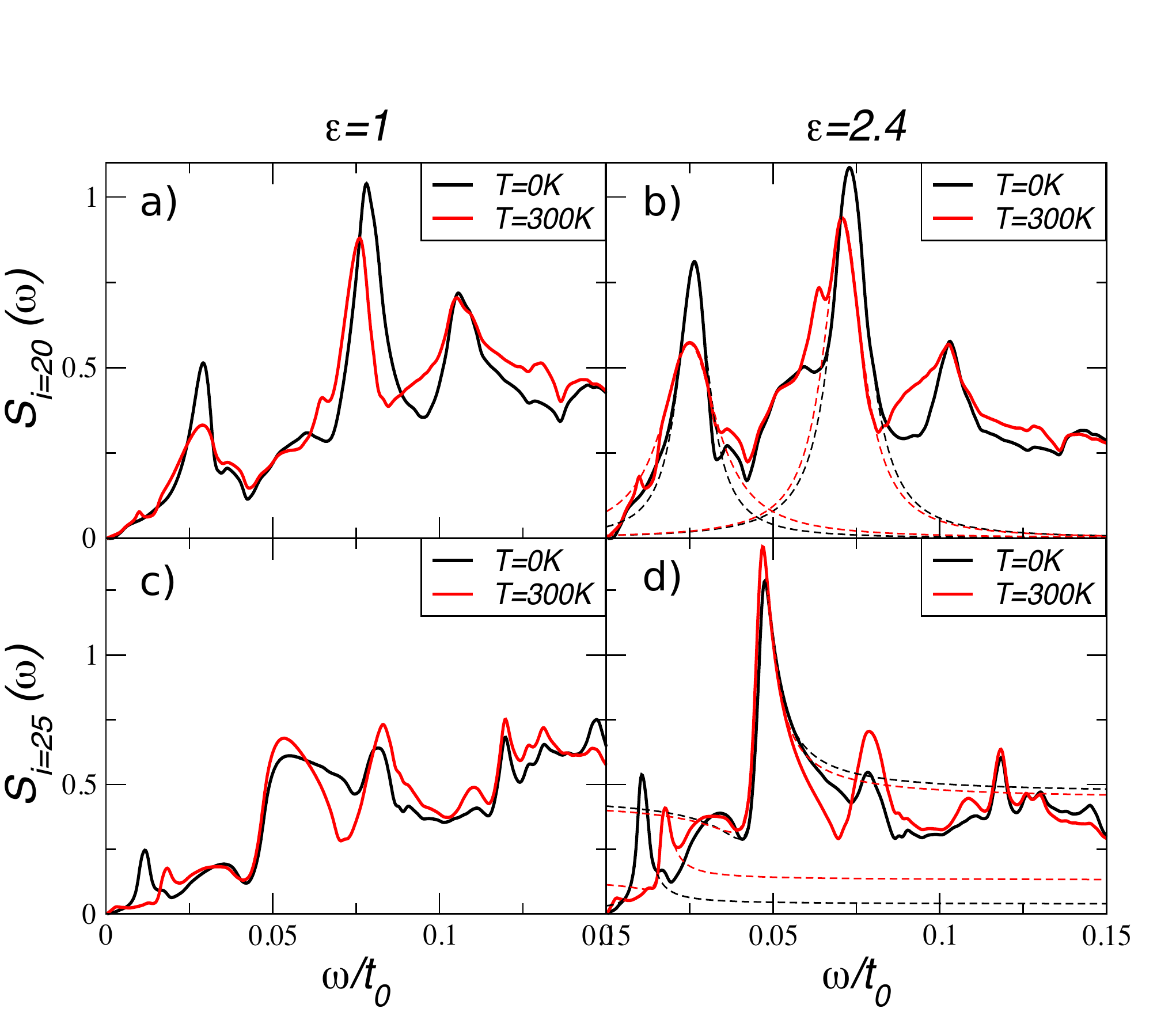}
\caption{\label{Fig_Plasmon_1} (Online color) Loss function $S_i(\q,\omega)=-$Im$\epsilon^{-1}(\q,\omega)$ for a twist angle $\theta_i$ a), b) with $i=20$ and c), d) $i=25$ for in-plane momentum $q=0.02a^{-1}$ as a function of frequency $\omega$ for two temperatures $T=0$ (black) and $T=300$K (red) as well as two effective dielectric media a), c) $\epsilon=1$ and b), d) and $\epsilon=2.4$. Fits to a b) Lorentzian and d) Fano resonance are shown as dashed curves.}
\end{figure}

In Fig. \ref{Fig_Plasmon_1}, line cuts at constant momentum $q= 0.02a^{-1}$ are shown for two temperatures $T=0$ (black) and $T=300$K (red). Panels a) and b) display three distinct peaks and for an effective dielectric medium with $\epsilon=1$ (suspended sample) the first two are related to a zero in the real part of the dielectric function, i.e., Re$\epsilon(\q,\omega_p)$=0 \cite{Footnote}. We can therefore speak of plasmonic excitations even though they are damped by a non-zero imaginary part. For a dielectric with $\epsilon=2.4$ (e.g., twisted bilayer on top of SiO${}_2$), only the first peak corresponds to a zero. Concerning the second and third peak, we associate them to transitions between the quasi-localised states inside the AA-stacked regions, see also the discussion below. The main resonances can be well fitted by a Lorentzian, see dashed lines in Fig. \ref{Fig_Plasmon_1} b). 

In Fig. \ref{Fig_Plasmon_1} c) and d), the same curves are plotted for a twist angle with $i=25$. Now, also for a dielectric substrate up to $\epsilon=4$, genuine plasmons with Re$\epsilon=0$ are present at $T=0$, whereas for suspended samples, most resonances are smeared out. Due to the stronger confinement, the two lowest resonances show an asymmetric line shape which can be well fitted by a Fano resonance, see dashed lines in Fig. \ref{Fig_Plasmon_1} d) .  

\begin{figure}[t]
\includegraphics[width=0.99\linewidth]{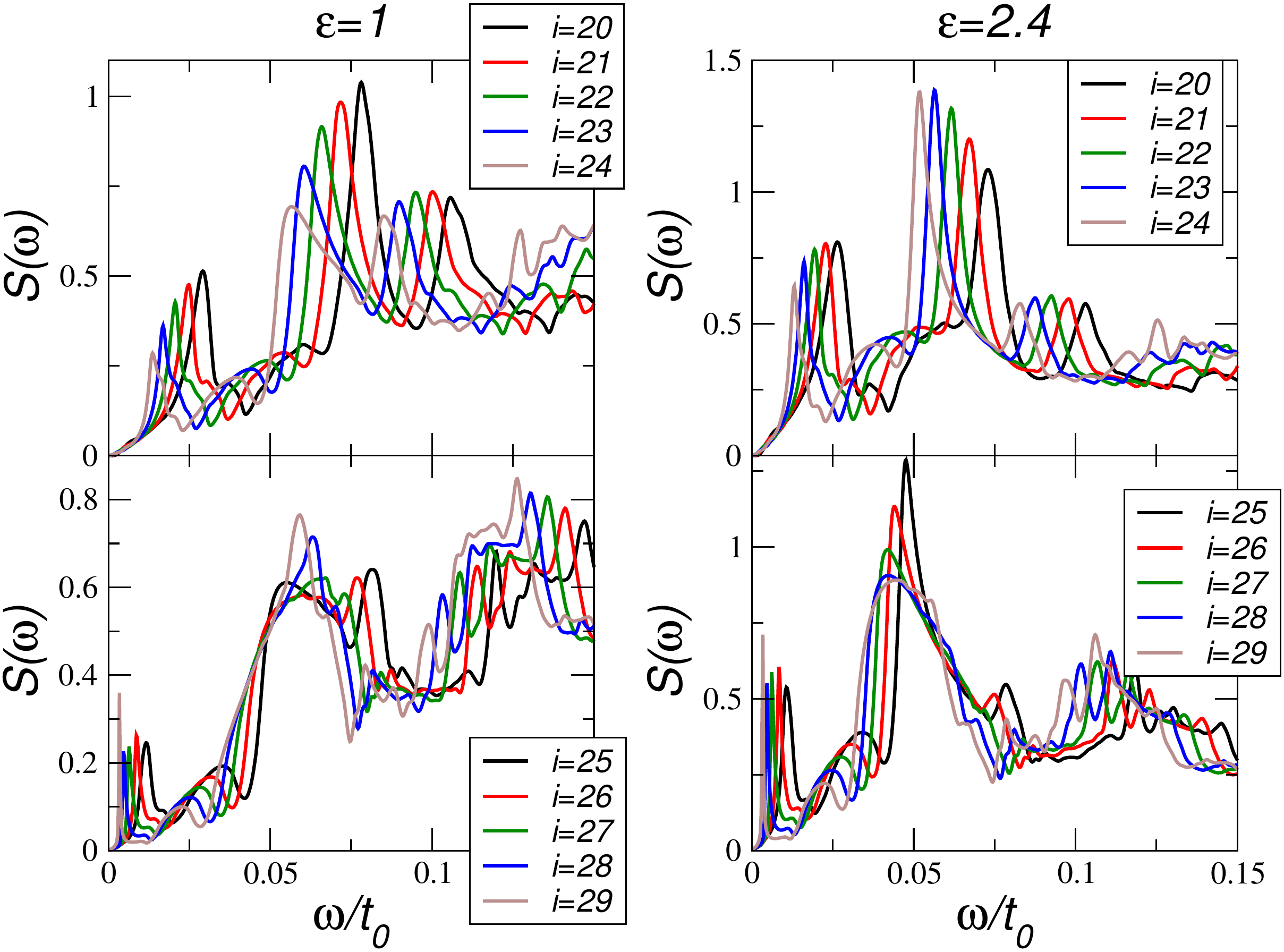}
\caption{\label{LossFunctionAngle} (Online color) Loss function $S_i(\q,\omega)=-$Im$\epsilon^{-1}(\q,\omega)$ for a twist angles $\theta_i$ with $i=20,21,22,23,24$ (upper panels) and $i=25,26,27,28,29$ (lower panels) and wave number $q=0.02a^{-1}$ as a function of frequency $\omega$ for zero temperature $T=0$ (black) and two effective dielectric media $\epsilon=1$ (left) and $\epsilon=2.4$ (right).}
\end{figure}

In Fig. \ref{LossFunctionAngle}, the loss function is shown for various angles at $T=0$ for two different effective dielectric media with $\epsilon=1$ (left) and $\epsilon=2.4$ (right). As can be seen, the loss function becomes more asymmetric for increasing $i$, i.e., for decreasing twist angle, consistent with our assumption that the plasmonic resonance can be described by Fano-resonances. We also see a linear shift in the resonant plasmon energies in accordance to a $1/R$-behaviour, where $R\sim\sqrt{A_i}\sim i$ denotes the radius of the localised region which is approximately linear for large $i\gtrsim15$.

In table \ref{table}, the fitting parameters for $i=20$ and $\epsilon=2.4$ at $qa=0.02$ are given and in table \ref{tableFano2}, for $i=25$ and $\epsilon=2.4$. From the quality factor $\omega_0/\gamma$, we see that there are damped oscillations of up to 10 cycles. The quality factor is even enhanced for larger momentum by a factor of 2--3.
\begin{table}[t]
\begin{center}
\begin{tabular}{c||c|c|c|c}
 &$m=1$&$m=1$&$m=2$&$m=2$\\
 &$T=0$K&$T=300$K&$T=0$K&$T=300$K\\
\hline
$\hbar\omega_0/t_0$&0.026&0.025&0.073&0.071\\
\hline
$\gamma/t_0$&0.0055&0.010&0.0065&0.070\\
\hline
$\hbar C/t_0^2$&0.81&0.570&1.08&0.94\\
\hline
$\omega_0/\gamma$&4.7&2.5&11.2&10.1 
\end{tabular}
\caption{Fitting parameters for the loss function approximated by the Lorentzian $f (\omega)=C [(\omega- \omega_0)^2+\gamma^2]^{-1}$.} 
  \label{table}
\end{center}
\end{table}
\begin{table}[t]
\begin{center}
\begin{tabular}{c||c|c|c|c}
 &$m=1$&$m=1$&$m=2$&$m=2$\\
 &$T=0$K&$T=300$K&$T=0$K&$T=300$K\\
\hline
$Q$&8&2.8&2.2&2.8\\
\hline
$\hbar\omega_0/t_0$&0.0105&0.0171&0.0458&0.0455\\
\hline
$\gamma/t_0$&0.002&0.0023&0.0029&0.0027\\
\hline
$\hbar S_{max}/t_0^2$&0.536&0.407&1.284&1.411\\
\hline
$\hbar S_{min}/t_0^2$&0.030&0.096&0.291&0.316\\
\hline
$\omega_0/\gamma$&5&7.4&16.4&16.9
\end{tabular}
\caption{Fitting parameters for the loss function approximated by the Fano resonance $f_Q(\omega)=S_{min}+\widetilde C (Q\gamma +\omega-\omega_0 )^2[(\omega- \omega_0)^2+\gamma ^2]^{-1}$ with $\widetilde C=(S_{max}-S_{min})/(1+Q^2)$.} 
  \label{tableFano2}
\end{center}
\end{table}
\section*{Local field effects}
For small twist angles and large momentum, local field effects have to be taken into account since the wave number $q$ becomes comparable to the length of the reciprocal lattice vector $|\G_i|=|\G_0|/\sqrt{A_i/3}$ with $|\G_0|=4\pi/(\sqrt{3}a)$ \cite{giu05}. Here, we will analyse the effect and show that it can be neglected. 

The response of a periodic structure to a plane wave with wave number $\bq$ is given by a Bloch wave which can be written as a superposition of plane waves with $\bq+\G$ where $\G$ are arbitrary reciprocal lattice vectors. The linear response thus reads $\rho_{\bq+\G}(\omega)=\sum_{\G}\chi_{\G,\G'}(\q,\omega)\phi_{\bq+\G'}(\omega)$ and the dynamical response matrix is given by 
\begin{widetext}
\begin{eqnarray}
\chi_{\G,\G'}(\q,\omega)&=&\frac{g_s}{V}\sum_{\tau=\pm} \sum_{m,n} \sum_{\bk\in1.BZ}\frac{n_F[E_m^\tau(\bk)]
 - n_F[E_n^\tau(\bk+\bq)]}{\hbar\omega- E_n^\tau(\bk+\bq)+E_m^\tau(\bk)+i0}\\
 &\times&\langle \tau,m,\k|e^{-i(\q+\G)\hat\r}|\tau,n,\k+\q\rangle\langle\tau,n,\k+\q|e^{i(\q+\G')\hat\r}|\tau,m,\k\rangle\;,\nonumber
\end{eqnarray}
\end{widetext}
where the notation follows Eq. (1) of the main text. As a consequence of the above expression, we have $\chi_{\G,\G'}(\bq,\omega)=\chi_{\G',\G}^*(\bq,\omega)$ and time-reversability demands $\chi_{\G,\G'}(\bq,\omega)=\chi_{\G',\G}(-\bq,\omega)$. Due to a real response, we further have $\chi_{\G,\G'}(\bq,\omega)=\chi_{-\G,-\G'}^*(-\bq,-\omega)$. 

The plasmonic modes can be discuss by the dielectric function which is given by the following matrix within the random-phase approximation:\begin{equation}
\epsilon_{\G,\G'}(\q,\omega)=\left[\delta_{\G,\G'}-v_\G(\q)\chi_{\G,\G'}(\q,\omega)\right]
\end{equation}
with $v_\G(\q)=v(\q+\G)=\frac{e^2}{2\epsilon_0\epsilon |\q+\G|}$. The loss function is then given by $S(\q,\omega)=-\im [\epsilon^{-1}]_{\G=0,\G'=0}(\q,\omega)$ \cite{giu05}.

\begin{figure}[h]
\includegraphics[width=0.99\linewidth]{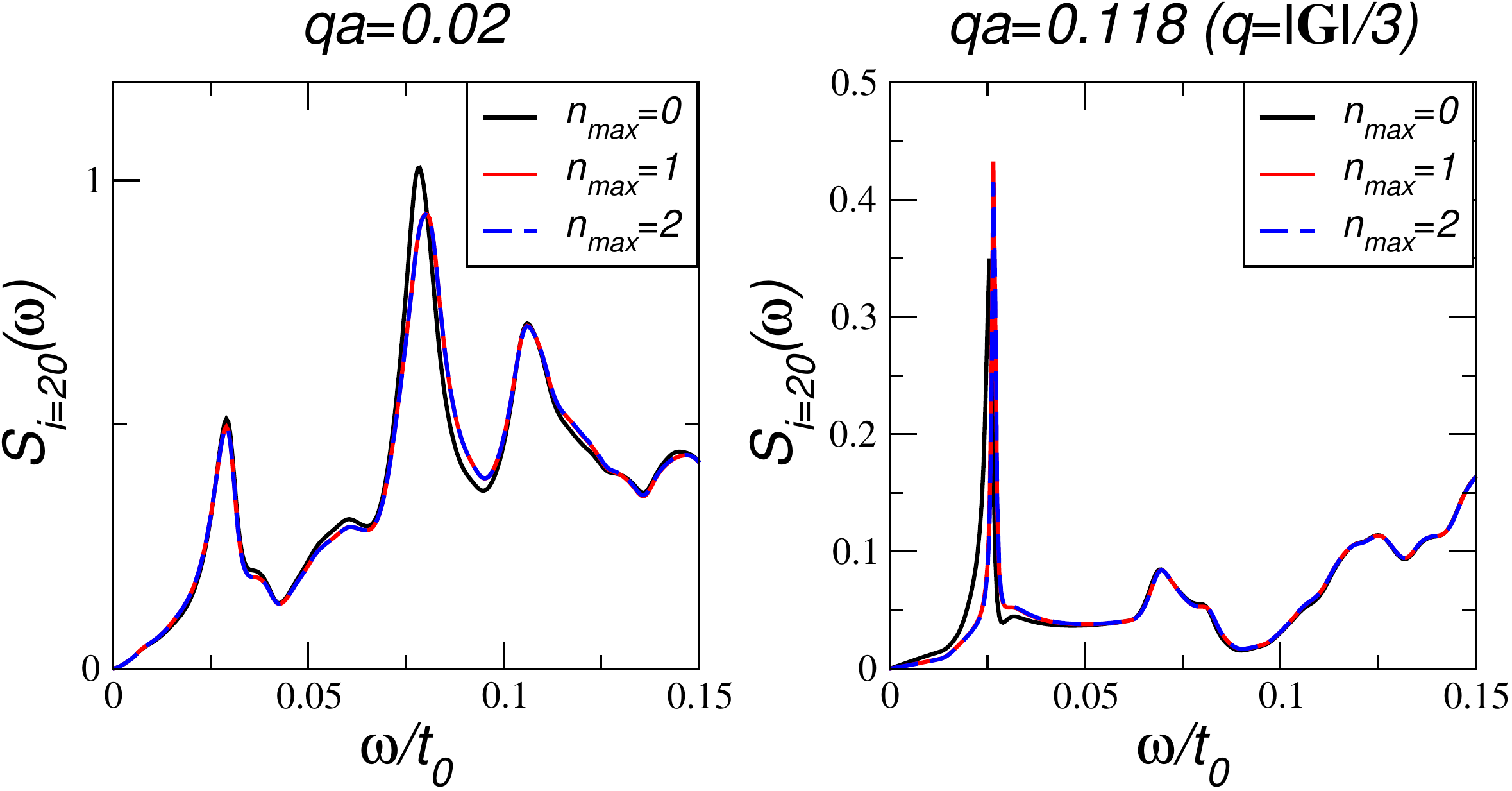}
\caption{\label{FigLocalFieldEffects} (Online color) Loss function $S_{i=20}(\q,\omega)=-\im\epsilon^{-1}_{0,0}(\q,\omega)$ for two momenta $q=0.02a^{-1}$ (left) and $q=0.118a^{-1}$ (right) at zero temperature $T=0$ and dielectric medium $\epsilon=1$. The direction of the momentum vector was chosen along the $KK'$-line and $q=|\G|/3$ connects the two $K$-points.}
\end{figure}

In Fig. \ref{FigLocalFieldEffects}, the loss function for twist angle $\theta_{i=20}$ is shown at two momenta $q=0.02a^{-1}$ (left) and $q=0.118a^{-1}$ (right). The reciprocal lattice vectors that build up the matrix are given by $\G = n_1\G_1+n_2\G_2$ with $-n_{max}\leq n_1,n_2\leq n_{max}$ and we show the results for $n_{max}=0,1,2$ corresponding to matrices of dimensions 1,9, and 25. The loss function with $n_{max}=1$ is almost identical to the one with $n_{max}=2$ and convergence is thus rapidly reached.  

For small momentum $q=0.02a^{-1}$, there are only small differences between the loss-function obtained from different matrices and we can thus neglect local field effects for this parameter regime. For large momentum $q=0.118a^{-1}$, the changes are slightly more pronounced and a shift of the first resonance is seen. Still, they are surprisingly small and the main resonance prevails which gives rise to a stable plasmonic excitations at constant energy. This shows that local field effects for twisted bilayers can be neglected for the considered parameter regime discussed in this work.

\section*{Finite Doping and Bias}
Here, we discuss the dependence of the loss function on finite doping $\mu$ and also on a finite bias $\Delta$ between the two layers. Changing the electrochemical potential between the two layers will open up a gap in Bernal-stacked graphene bilayer, but for twisted bilayer graphene only the two Dirac points of the same valley are shifted to positive and negative energies, respectively. Nevertheless, a local gap in the AB-stacked regions for small twist angles is expected which can be characterised by the local density of states (LDOS). In order to discuss the finite bias dependence, we add the following term to the Hamiltonian of Eq. (\ref{ham1}):
\begin{equation}
\label{ham1Delta}
 \hat{H}_\Delta(\bk)= \Delta \sum_{\sigma=A,B}\left( \hat{a}^\dagger_{\phi=0,\sigma \bk}\hat{a}_{\phi=0,\sigma \bk}-\hat{a}^\dagger_{\phi=\theta_i,\sigma \bk}\hat{a}_{\phi=\theta_i,\sigma \bk}\right)
\end{equation} 
\begin{figure*}[t]
\includegraphics[width=0.32\linewidth]{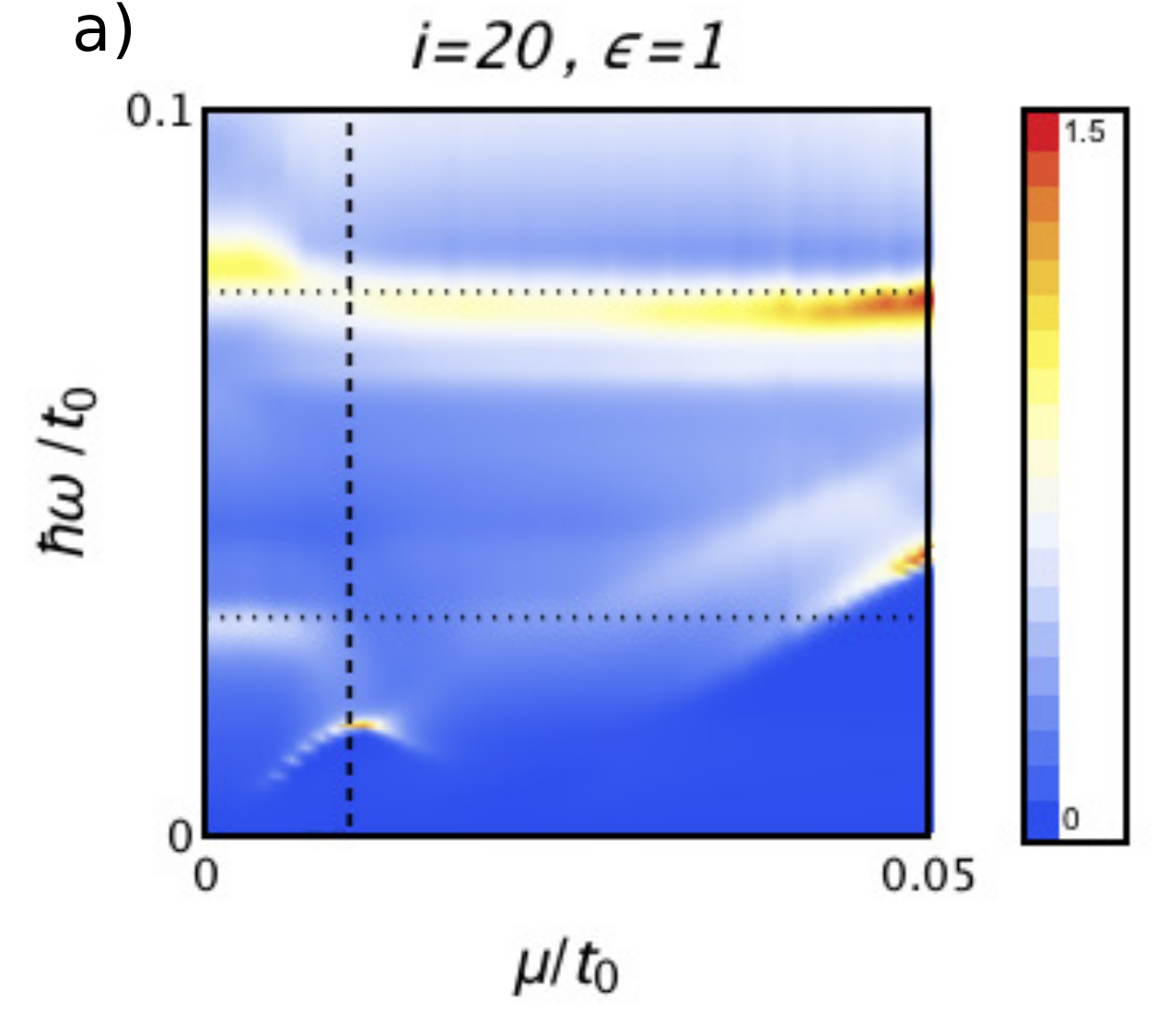}
\includegraphics[width=0.32\linewidth]{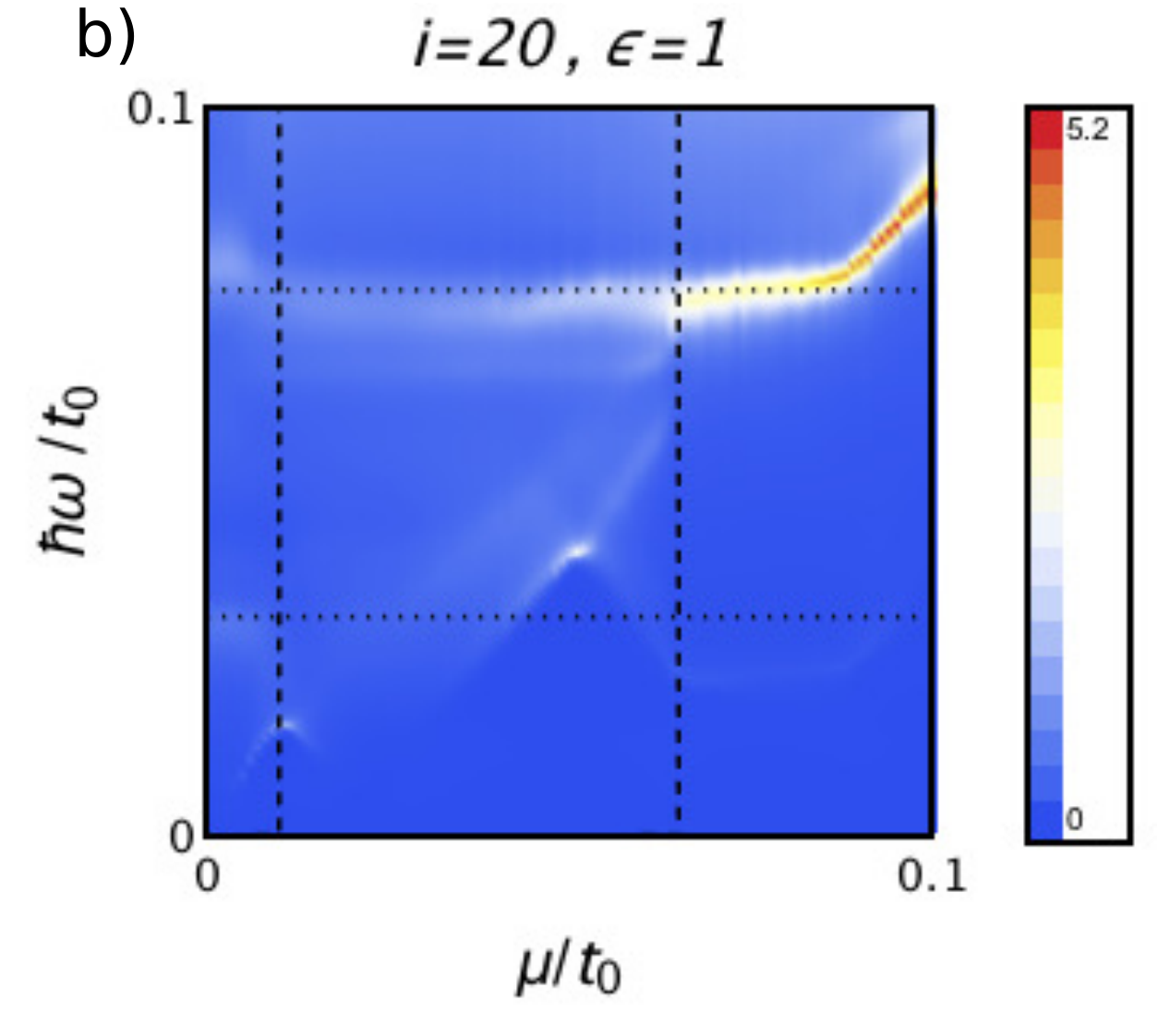}
\includegraphics[width=0.32\linewidth]{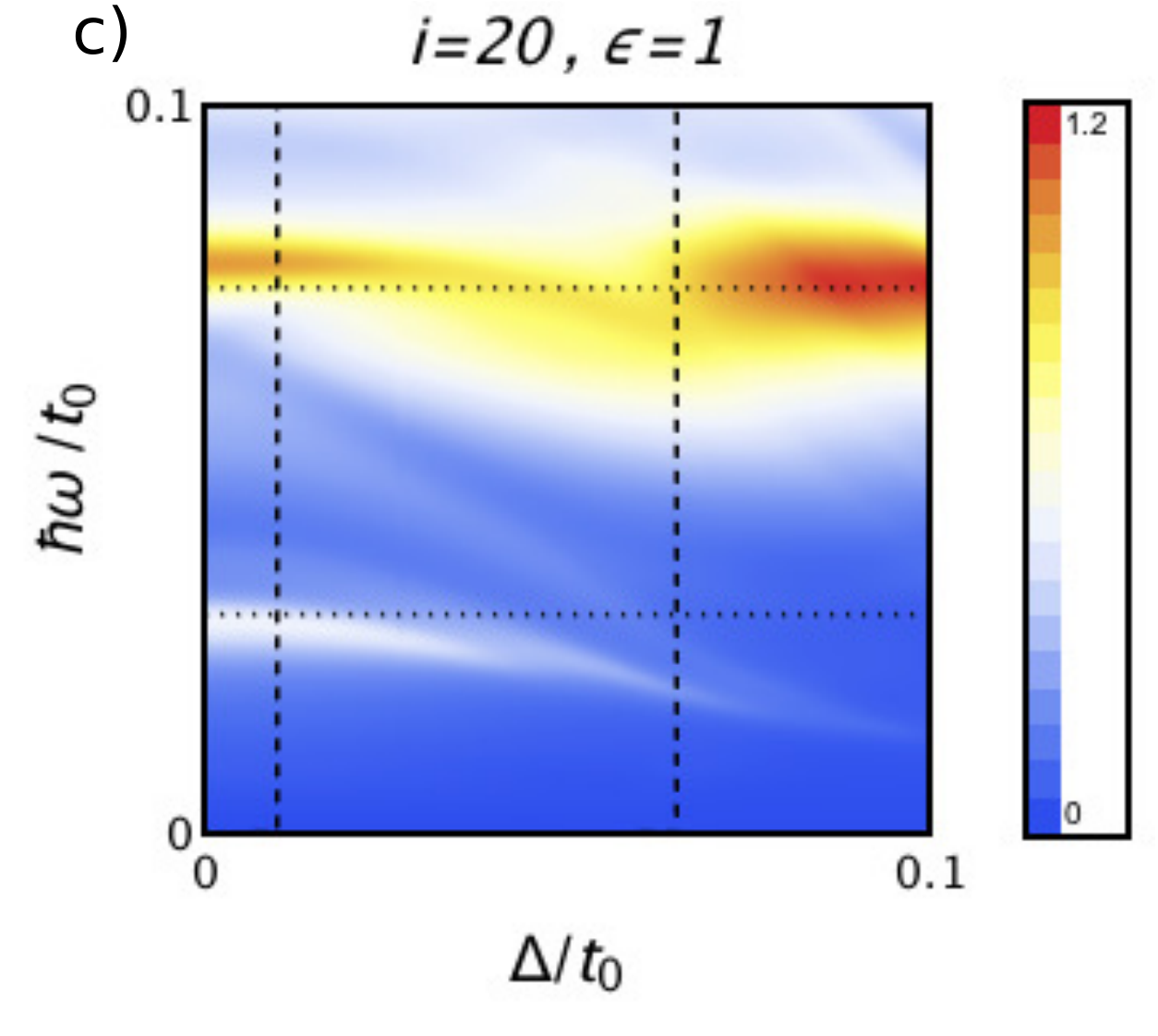}
\caption{\label{FiniteMu} (Online color) a) and b): Loss function $S(\q,\omega)=\im\epsilon^{-1}(\q,\omega)$ as function of the chemical potential $\mu$  on different scales. c) Loss function as function of the interlayer bias $\Delta$. The results are obtained for twist angle $\theta_{i=20}$ and fixed wave number $qa=0.02$ with the direction of the momentum vector along the $KK'$-line. Also shown vertical dashed lines at $\mu/t_0(\Delta/t_0)=0.01,0.065$ and horizontal dotted lines at $\hbar\omega/t_0=0.03,0.075$ as guide for the eye.}
\end{figure*}

In Fig. \ref{FiniteMu}, the loss function $S(\q,\omega)=\im\epsilon^{-1}(\q,\omega)$ for a twist angle with $i=20$ is shown as function of the chemical potential $\mu$ at different scales, see  a) and b). Also the loss function as function of the interlayer bias $\Delta$ can be seen in panel c). The momentum is fixed at $qa=0.02$ with direction parallel to $\Delta\bK$. In both cases, i.e., for the two variables $\mu$ and $\Delta$, we see that the lowest two resonances at $\hbar\omega_p^1=0.03$eV as well as at $\hbar\omega_p^2=0.075$eV persist for  $\mu,\Delta\lesssim\hbar\omega_p^m$ with $m=1,2$, respectively. They are then modified and we infer that the plasmonic modes are induced via interband transitions and that the plasmonic resonances can be quenched and enhanced by electrostatical gating which could be used as an optical switch. 

We also observe an enhanced resonance for finite interlayer bias with $\Delta\gtrsim\hbar\omega_p^m$. This can be understood from the local gap induced in the AB- and BA-stacked regions which surround the AA-stacked islands. This local gap will favor the localisation inside the AA-stacked islands and thus stabilise the plasmonic modes . 

\section*{Real space interpretation}
In this section, we want to explore a simple quantum dot model that can account for quasi-localised states in the AA-stacked islands. We will show that quasi-bound states exist for this model even when the confinement is related to a rather small energy scale, i.e., the interlayer hopping amplitude. The resulting spectrum of the bound-state shows is equidistant in energy and compares favourable to the spectrum of the loss function. Nevertheless, we wish to stress that our analysis can only motivate the resulting spectrum of the loss function.

For small angles, twisted bilayer graphene can be viewed as plackets of AA-stacked graphene surrounded by three AB- and BA-stacked regions, respectively. The band structure of AA-stacked graphene is simply the band structure of single layer graphene shifted to positive and negative energy $t_\perp$, respectively. There are no transitions allowed between the two conical bands and we can thus approximate the AA-stacked region by a simple graphene monolayer, $H_{sg}$, but with finite chemical potential $\mu=t_\perp$. 

Bernal (or AB) stacked graphene displays four parabolic branches also separated by $t_\perp$. The easiest way to model the interface between AA- and AB-stacked regions is thus given by a finite mass-term of approximate energy $t_\perp$. For sufficiently localised states, hybridisation of the wave functions of adjacent dots can be neglected, and it is enough to only consider one quasi-localised circular dot. Our simple model then reads
\begin{equation}
\label{DotModel}
H_{AA}^{island}=H_{sg}+t_\perp\sigma_z\theta(r-R_1)\theta(R_2-r)\;.
\end{equation}
We choose $R_1=7$nm and $R_2=8$nm in the case of a twist angle $\theta_{i=25}\approx1.30^\circ$ for which the length of the vector of the unit cell is $\sim11$nm. For generalisations to general periodic systems and wave function matching in graphene systems with circular symmetry including a mass-term, we refer to Ref. \cite{Gutierrez15}. 

\begin{figure}[h]
\includegraphics[width=0.99\linewidth]{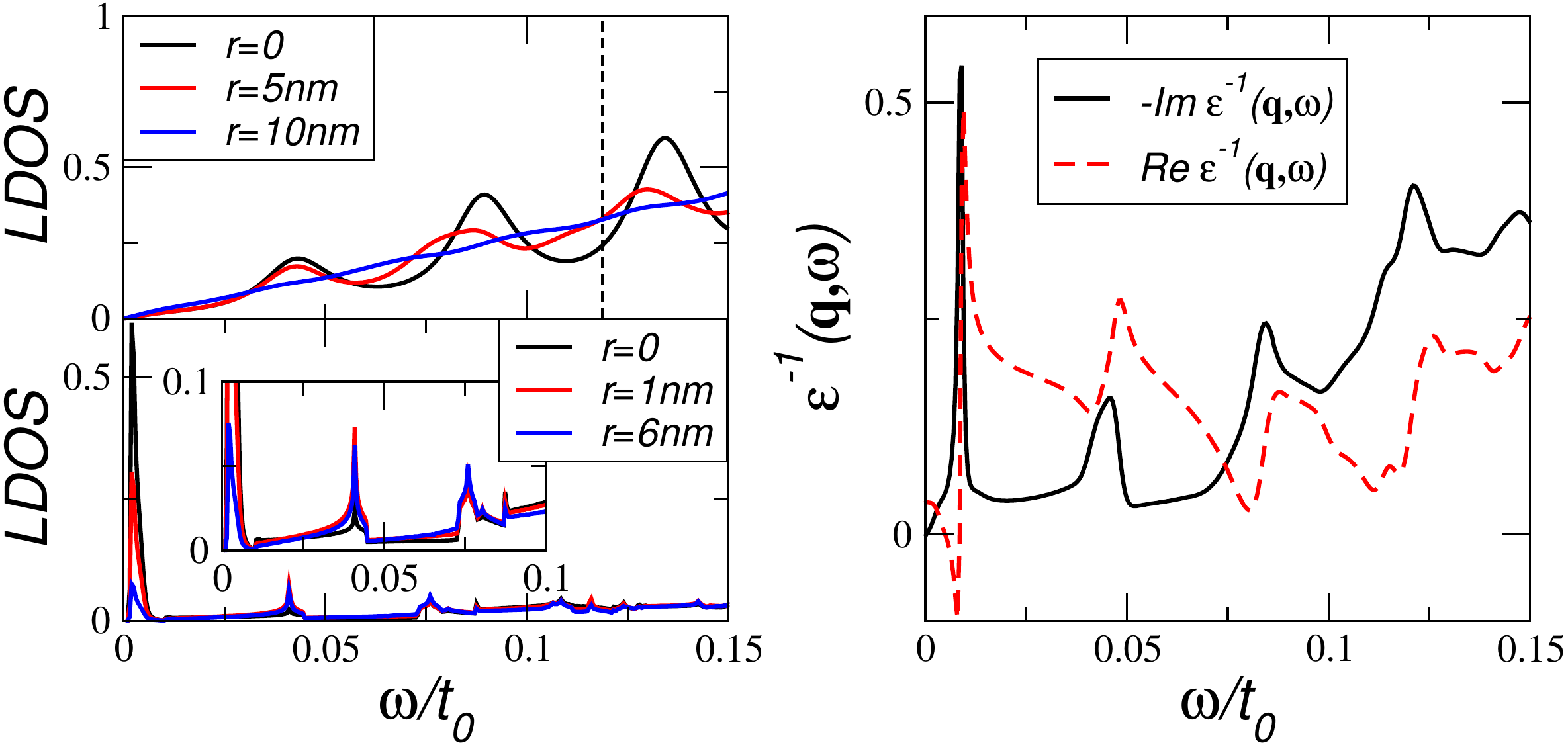}
\caption{\label{LDOS} (Online color) Left hand side: Upper panel: Local density of states (LDOS) of Eq. (\ref{DotModel}) for three different positions $r=0, 5,10$nm for a dot with radius $R_1=7$nm and mass confinement with barrier height $\Delta=t_\perp$ between $R_1$ and $R_2=8$nm, see SI. Lower panel: LDOS of Eq. (\ref{ham1}) with $\theta_{i=25}$ for $r=0,1,6$nm. The inset highlights the localised states. Right hand side: The imaginary (loss function) and real part of the inverse dielectric function $\epsilon^{-1}(\q,\omega)$ for twisted bilayer graphene with $\theta_{i=25}$ at $qa=0.1$ (direction along the KK'-line).}
\end{figure}

In Fig. \ref{LDOS}, we compare the local density of states (LDOS) for the simple dot model, Eq. (\ref{DotModel}), and the full continuous model, Eq. (\ref{ham1}), with the loss function of twisted bilayer graphene with $\theta_{i=25}$ at $qa=0.1$. On the left, the LDOS is shown for three different positions (center, edge, and outside the dot confinement) and the chemical potential at $\mu=t_\perp$ is indicated by the dashed vertical line in the upper panel. The inset of the lower panel highlights the localised states. On the right, we observe that the first peak in the loss function is related to the zero of the real part of $\epsilon(\q,\omega)$ and thus resembles a genuine plasmonic resonance. The other peaks can be related by inter/intraband transition between the localised levels inside the dot, i.e, the peaks of the LDOS are separated by $\Delta E=0.45t_0$, similar to the energy separation of the loss function. 

\begin{thebibliography}{54}
\bibitem [{Foo()}]{Footnote}%
  \BibitemOpen
\bibfield  {journal} {  }\href@noop {} {}\bibinfo {note} {We note that only
  every second zero is related to a peak in the loss function due to the
  structure of the imaginary part of the response function.}\BibitemShut
  {Stop}%
\end{thebibliography}
\end{document}